\documentclass[twocolumn,showpacs,amsmath,nofootinbib,pra,aps,amssymb,longbibliography]{revtex4-2}
\usepackage[T1]{fontenc}
\usepackage[utf8]{inputenc}
\usepackage{color} 
 
\usepackage{times}
\usepackage{array}
\usepackage{amssymb,amsmath}
\usepackage{amsbsy}
\usepackage[pdftex]{graphicx}
\usepackage{float}
\usepackage{dcolumn}
\usepackage{scalerel}
\usepackage{tikz}

\usepackage[unicode,breaklinks]{hyperref}
\hypersetup{
    unicode=true,
    plainpages=false, 
    colorlinks=true,
    linkcolor=blue,
    citecolor=blue,
    filecolor=black,
    urlcolor=blue
}
\usetikzlibrary{svg.path}

\definecolor{orcidlogocol}{HTML}{A6CE39}
\tikzset{
  orcidlogo/.pic={
    \fill[orcidlogocol] svg{M256,128c0,70.7-57.3,128-128,128C57.3,256,0,198.7,0,128C0,57.3,57.3,0,128,0C198.7,0,256,57.3,256,128z};
    \fill[white] svg{M86.3,186.2H70.9V79.1h15.4v48.4V186.2z}
                 svg{M108.9,79.1h41.6c39.6,0,57,28.3,57,53.6c0,27.5-21.5,53.6-56.8,53.6h-41.8V79.1z M124.3,172.4h24.5c34.9,0,42.9-26.5,42.9-39.7c0-21.5-13.7-39.7-43.7-39.7h-23.7V172.4z}
                 svg{M88.7,56.8c0,5.5-4.5,10.1-10.1,10.1c-5.6,0-10.1-4.6-10.1-10.1c0-5.6,4.5-10.1,10.1-10.1C84.2,46.7,88.7,51.3,88.7,56.8z};
  }
}

\newcommand{\orcidicon}[1]{\href{https://orcid.org/#1}{\mbox{\scalerel*{
\begin{tikzpicture}[yscale=-1,transform shape]
\pic{orcidlogo};
\end{tikzpicture}
}{|}}}}

\usepackage{url}
\usepackage{verbatim}

\begin{document}
\title{Supersolid spectroscopy}
\author{L.~M.~Platt\orcidicon{0009-0002-2590-3153}}
\author{D.~Baillie\orcidicon{0000-0002-8194-7612}}
\author{P.~B.~Blakie\orcidicon{0000-0003-4772-6514}}
\affiliation{$^{1}$Dodd-Walls Centre for Photonic and Quantum Technologies, Dunedin
9054, New Zealand~\\
 $^{2}$Department of Physics, University of Otago, Dunedin 9016,
New Zealand}
\date{\today}

\begin{abstract}
We develop a linear response theory to provide a unified description of two recent spectroscopy protocols for probing one-dimensional supersolid states realized in cold-atom systems. 
Both protocols involve applying a periodic optical potential to excite the supersolid and determine its excitation frequencies and density response characteristics. This information can  be used to estimate the superfluid fraction. We validate our linear response theory against nonlinear meanfield simulations of the dynamics for both translationally invariant and trapped cases. A key focus is the behavior at the band edge - the regime occurring when the optical potential used to excite the system has a wavelength that is twice the value of the supersolid lattice constant. Here symmetry can be used to selectively excite a mode from one of the two low-energy gapless excitation bands.  Finally, we consider the application of the spectroscopy protocols to determine the superfluid fraction, showing the relationship to hydrodynamic theory and a Josephson-Junction array model.
\end{abstract}
\maketitle

\section{Introduction} 

Recently  \ifmmode \check{S}\else Š\fi{}indik \textit{et
al.}~\cite{Sindik2024a} proposed a protocol for probing a supersolid state of a dipolar Bose Einstein condensate (BEC)
by abruptly removing an applied spatially periodic potential. The
resulting oscillations of the supersolid exhibited two frequency components that they related
to the quasiparticle excitation energy and static density response function
of the lowest two bands at the wavelength set by the perturbation.
For long-wavelength perturbations the lowest two excitation bands
are well-described by hydrodynamic theory, and this
protocol can be used to determine the compressibility, the elastic
modulus of the lattice, and the superfluidity.

Another protocol has been developed and applied to an experiment
with a dipolar supersolid by Biagioni \textit{et al.}~\cite{Biagioni2024a}.
This involved the brief application of a strong periodic potential, with a period of two lattice sites, 
to imprint a differential phase between adjacent sites. The subsequent dynamics
revealed a Josephson Junction-like oscillation involving the phase
difference and atom number difference between adjacent sites. This protocol was
used to provide direct evidence of a sub-unity superfluid fraction.

For brevity we will refer to the first approach as the density protocol
and the second as the phase protocol. Since both approaches involve
the application of a periodic spatially modulated potential, it seems
natural to expect that both approaches should be described within
a single framework. Also, some immediate questions emerge, such as:
Why does the density protocol excite oscillations with two frequency
components while the phase protocol excites only
a single frequency? Superfluidity is inherently a long-wavelength
property of the system, so how can it be determined by the phase protocol
with a relatively short-wavelength excitation?

\begin{figure}[htbp]
 \includegraphics[width=3.2in]{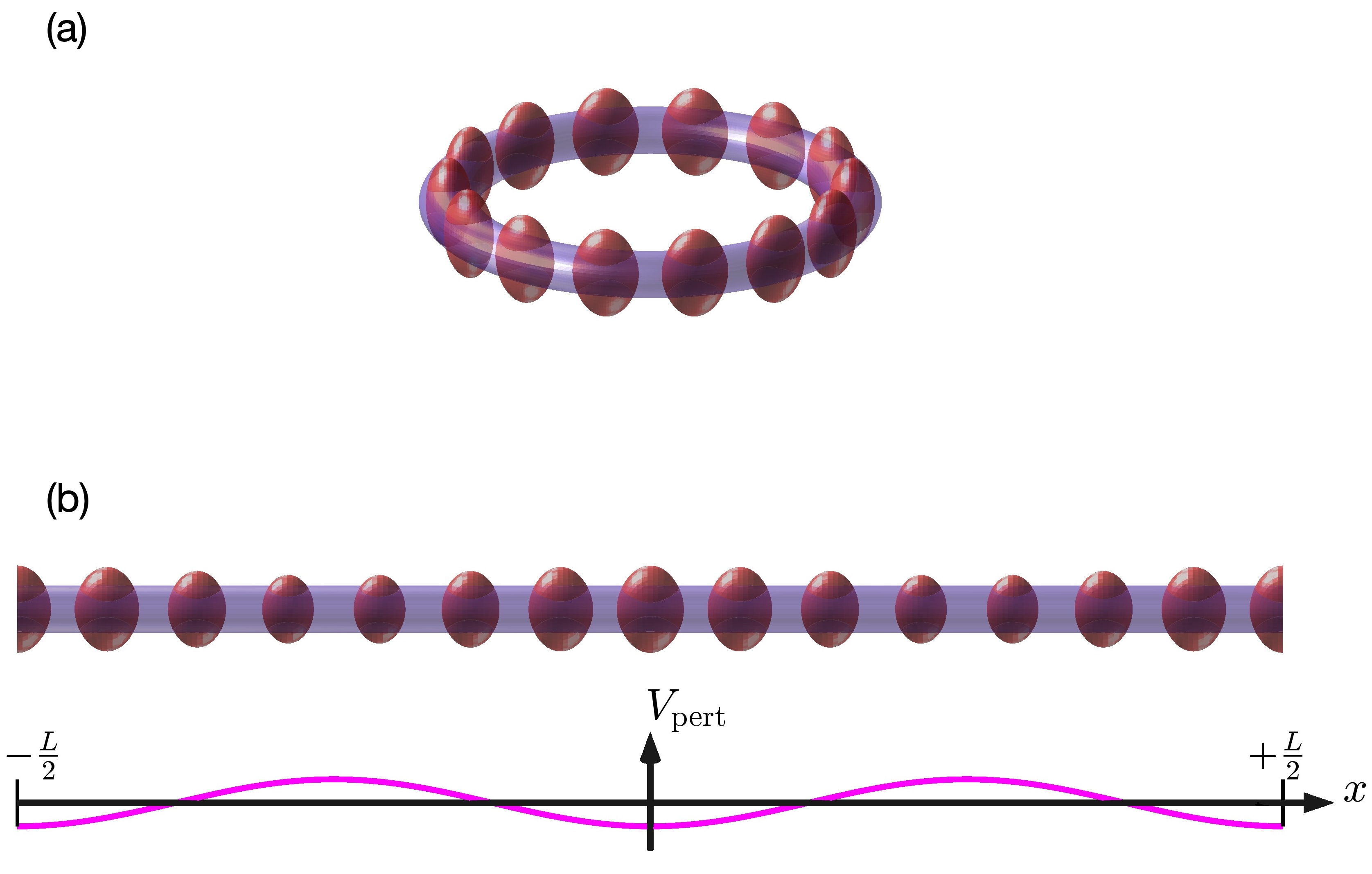}
 \caption{Schematic image of system and perturbation. (a) A supersolid (red) in a ring-shaped trap (blue).   (b) Equivalent (unwrapped) finite tube system of length $L$. Applied perturbation potential  $V_\mathrm{pert}$ shown for reference. }\label{figSch}
\end{figure}

In this paper we develop a general
linear response theory for probing supersolid states with one-dimensional
crystal structure. This is most transparently developed for the translationally
invariant case, which can be realized in experiments with a dipolar Bose-Einstein
condensate (BEC) supersolid \cite{Tanzi2019a,Bottcher2019a,Natale2019a} confined in a ring geometry [see Fig.~\ref{figSch}(a)]. This system and geometry has been the subject of several recent studies
\cite{Tengstrand2023a,Sindik2024a,Hertkorn2024a}. Neglecting curvature
effects, the ring system is equivalent to a finite interval of length
 $L$ (corresponding to the ring circumference) and subject to periodic boundary conditions
[Fig.~\ref{figSch}(b)]. Previous work on the ground states and
excitations of a purely linear geometry tube confined dipolar BEC
\cite{Roccuzzo2019a,Smith2023a,Blakie2023a} are in good qualitative
agreement with calculations performed in the ring potential \cite{Sindik2024a}.

In this work we will mainly illustrate our results using a soft-core
model of a supersolid.  This is purely a 1D supersolid model, with
the advantage that the calculations are relatively straightforward compared to the dipolar tube model, which involves full 3D calculations and
some intricate technical details of dealing with the singular dipole-dipole interactions.
However, for probing along the supersolid [i.e.~along the $x$-axis in Fig.~\ref{figSch}(b)], and at sufficiently low energies that transverse excited states can be ignored, the 1D
soft-core and tube dipolar systems behave similarly (e.g.~see the
comparisons made in Ref.~\cite{Platt2024a}). Notably, both models
have a continuous superfluid to supersolid phase transition, and exhibit
two gapless excitation branches.

The outline of this paper is as follows. In Sec.~\ref{Sec:Protocols} we introduce the density and phase protocols  along with a brief description of the soft-core model. The system dynamics obtained by solving the Gross-Pitaevskii (GP) equation are investigated, and the case of probing at the band-edge is discussed. We also propose generalized density and phase protocols, which removes the assumption that the  perturbation and observable we probe are aligned to the supersolid. In Sec.~\ref{Sec:LRT} we present the linear response theory for the density and phase protocols, and the generalized protocols. We use this to understand the dynamics seen in Sec.~\ref{Sec:Protocols}.
In Sec.~\ref{Sec:Trap} we present results for a supersolid with confinement along the $x$-axis to assess its effect on system behavior compared to the translationally invariant case.
We focus on the question of superfluidity in  Sec.~\ref{Sec:SF}, discussing the relationship to hydrodynamic quantities and a Josephson-Junction array model. This allows us to make some comments on the suitability of the two protocols for determining superfluidity.  Finally, we summarise and give concluding remarks in Sec.~\ref{Sec:Conclusions}.

\section{Spectroscopy protocols}\label{Sec:Protocols}
\subsection{System and perturbation}\label{Sec:SysPert}
Consider a dilute BEC in a supersolid state
confined in a ring geometry [Fig.~\ref{figSch}(a), also see  \cite{Tengstrand2023a,Sindik2024a,Hertkorn2024a}]. We take the supersolid to have $M_s$ lattice sites, and restrict our focus to the case where $M_s$ is even\footnote{This restriction is necessary to examine the proposal of Ref.~\cite{Biagioni2024a}, where an alternating phase is written on adjacent lattice sites.}. Both spectroscopy protocols we examine involve the application of the perturbation potential to the supersolid of the form
$V_{\mathrm{pert}}=-V_{0}\cos(M\phi)$, where $V_0$ is the amplitude of the potential, $\phi$ is the azimuthal
angle around the ring and $M$ is a positive integer.  
For large ring diameters we can neglect curvature effects and map the system to a tube
of length $L$, corresponding to the ring
circumference  [see Fig.~\ref{figSch}(b)], and impose periodic boundary conditions on the tube.
Here we introduce the $x$-axis as going around the tube axis with
domain $-\frac{1}{2}L<x\le\frac{1}{2}L$. The wavevector $k=2\pi M/L$
describes the perturbation periodicity, i.e.~$V_{\mathrm{pert}}=-V_{0}\cos(kx)$.  

For a supersolid with lattice constant $a$, we will be interested in perturbation wavevectors $k$ in the range ${2\pi/L}\le k\le Q$, where   $Q\equiv \pi/a$ is the band-edge wavevector (i.e.~half the reciprocal lattice vector).  
The lower limit being a single variation around the ring, which takes the limit
$k\to0$ for large $L$, accessing the long-wavelength excitations of the system. The upper limit corresponds to a periodic variation occurring over two sites.   We avoid   $k$ being 0 or an integer multiple of the reciprocal lattice vector, as this would couple to zero energy excitations and linear response theory does not apply.
The theory we develop here is more generally applicable to  $k>Q$, but for wavevectors in our specified range, the perturbation most strongly couples to the lowest two gapless excitation bands, which are most sensitive to the manybody physics of the supersolid (cf.~higher energy Bragg spectroscopy in Refs.~\cite{Petter2021a,Chomaz2020a}).

To demonstrate the spectroscopy protocols and develop a linear response theory, we consider a straightforward supersolid model: a purely one-dimensional soft-core BEC. Because the low energy excitations of a 1D supersolid are universal in character, i.e.~two gapless bands, and weak perturbations do not couple to the transverse excitations in a tube confined dipolar supersolid, the 1D soft-core  system  provides an appropriate platform for developing our theory with the general result being immediately applicable to the tube-dipolar case. For reference we note the comparison of a 3D tube dipolar supersolid to a 1D soft-core system in Ref.~\cite{Platt2024a}, which demonstrates the general similarities of these models.

We consider a BEC of $N_T$ atoms on 1D domain
of length $L$, with periodic boundary conditions. The atoms interact with a soft-core  potential $U_{\mathrm{sc}}(x)=U_{0}\theta_H(a_{\mathrm{sc}}-|x|)$,
where  $\theta_H$ is the Heaviside step function, $a_{\mathrm{sc}}$ is the core radius and $U_{0}$ is the potential
strength. The meanfield description of this system is provided by the time-dependent GP equation $i\hbar \dot{\psi} =\mathcal{L}\psi$, where
\begin{align}
\mathcal{L}=-\frac{\hbar^2}{2m}\frac{d^2 }{d x^2} +V_{\mathrm{pert}} +\int dx^\prime\,U_{\mathrm{sc}}(x-x^\prime)|\psi(x^\prime)|^2,\label{LGP}
\end{align}
is the GP operator, and $V_{\mathrm{pert}}$ represents the applied perturbation. 
It is conventional to define the dimensionless interaction
parameter 
\begin{align}
\Lambda=\frac{2ma_{\mathrm{sc}}^{3}U_{0}N_T}{\hbar^{2}L}.\label{Lambda}
\end{align}
 A continuous transition occurs from a uniform to a modulated
state at the critical value $\Lambda_{c}=21.05$.  
Details about the excitations of this model are given in Sec.~\ref{Sec:excitations} (also see Refs.~\cite{Biagioni2024a,Kunimi2012a,Prestipino2018a}).

In solving for the supersolid ground states we choose for there to be a density peak (i.e.~lattice site) at $x=0$. This imposes an alignment with the potential [recall $V_\mathrm{pert}=-V_0\cos(kx)$] in that it has a trough at $x=0$ [like the situation in Fig.~\ref{figSch}(b), also see example ground state in Fig.~\ref{figEdgeDSF}(b)]. This alignment is explicit and necessary in the scheme of Ref.~\cite{Biagioni2024a}. We revisit this assumption in Sec.~\ref{Sec:GenP} where we introduce the generalized protocols and allow the perturbation (and observables) to be offset relative to the supersolid.

\subsection{Density protocol}\label{Sec:DP} 
  
 We first describe the \ifmmode \check{S}\else Š\fi{}indik \textit{et
al.}~\cite{Sindik2024a} proposal. Here the perturbation is considered to have been on for a long time such that the supersolid is in the ground state of the static perturbation. The perturbation strength is then suddenly set to zero at time $t=0$, and the dynamics is examined. For this protocol  the perturbation potential [i.e.~perturbation appearing in Eq.~(\ref{LGP})] has the form
\begin{align}
V_{\mathrm{pert}}\to{V}_d(x,t)=-V_0\theta_H(-t)\,\cos(kx).\label{VpDensity}
\end{align}
At $t>0$ the translational invariance around the ring is restored by the sudden removal of the potential, however this causes longitudinal phonon modes to propagate through the supersolid. 
To quantify the excitation  \ifmmode \check{S}\else Š\fi{}indik \textit{et
al.}~\cite{Sindik2024a} proposed measuring the linear density weighted by a cosine at wavevector $k$, i.e.,~the observable 
\begin{align}
{F}(t)=\int dx\,\cos(kx)|\psi(x,t)|^2,\label{Fobs}
\end{align}
and considered the evolution of $\delta F(t)\equiv F(t)-F_0$, where $F_0$ is the observable evaluated with the unperturbed supersolid ground state\footnote{Unless $k$ is equal to a reciprocal lattice vector $F_0=0$.}.

In Figs.~\ref{figLRdynamics}(a) and (b) we show the dynamics of
a soft-core supersolid state following the density protocol
outlined above.  In Fig.~\ref{figLRdynamics}(a) we compare the  dynamics of $\delta F(t)$ obtained from a GP simulation\footnote{Initial condition ($t=0$) is the ground state with the static perturbation.} for various values of probe strength $V_{0}$. These results show that for sufficiently small perturbation strengths  ($V_{0}\lesssim\hbar\omega_{0}$) the scaled observable $\delta F(t)/V_0$ is independent of $V_0$, indicating that the system is in the linear response regime.  Here we will focus on describing the behavior in the linear response regime, which we will later relate to system properties via linear response theory. When the system is probed with wave vector $k=Q/4$, $\delta F(t)$ clearly oscillates with two frequencies [i.e.~Fig.~\ref{figLRdynamics}(a)]. Similar results are obtained for any $k<Q$. This was the generic type of behavior explored in Ref.~\cite{Sindik2024a}, who proposed fitting the response to two cosines to extract the properties of the two lowest excitation branches of the supersolid.
 In contrast, in Fig.~\ref{figLRdynamics}(b), where the probe is at the band edge value $k=Q$, the response signal has a single dominant frequency.

\subsection{Phase protocol}\label{Sec:PP}

In Biagioni \textit{et al.}~\cite{Biagioni2024a} an alternative spectroscopy approach was suggested  involving the application of a perturbation as a short and strong pulse to the supersolid ground state to imprint a phase profile. While this protocol was presented for the case $k=Q$,  here we generalize it to any wavevector
$k$. Taking the idealization of a delta-function pulse, the perturbation potential is
\begin{align}
V_{\mathrm{pert}}\to V_{p}(x,t)=-V_0\delta t\delta(t)\,\cos(kx),\label{VpPhase}
\end{align}
with dimensionless pulse area $V_{0}\delta t/\hbar$. Immediately
following the pulse ($t\to0^{+}$) the wavefunction corresponds to the ground state  of the unperturbed system, $\psi_0$, with a phase written on it, i.e.~
\begin{align}
\psi(x,t=0^{+})=\psi_{0}(x)e^{iV_{0}\delta t\cos(kx)/\hbar}.\label{psiPhase}
\end{align}
Biagioni \textit{et al.}~considered the population and phase difference between adjacent sites as the relevant observables [see Sec.~\ref{Sec:JJA}]. When there are strong connections between supersolid sites (i.e.~when the supersolid does not consist of well-isolated droplets), defining the site population and phase is somewhat  arbitrary.  
Here we instead consider the $F$ observable (\ref{Fobs}), noting that for  $k=Q$,  $F$ corresponds to a weighted population difference of adjacent sites [i.e.~the $\cos(Qx)$ factor positively (negatively) weights the population at even (odd) sites in this observable].

\begin{figure}[htbp]  \includegraphics[width=3.2in]{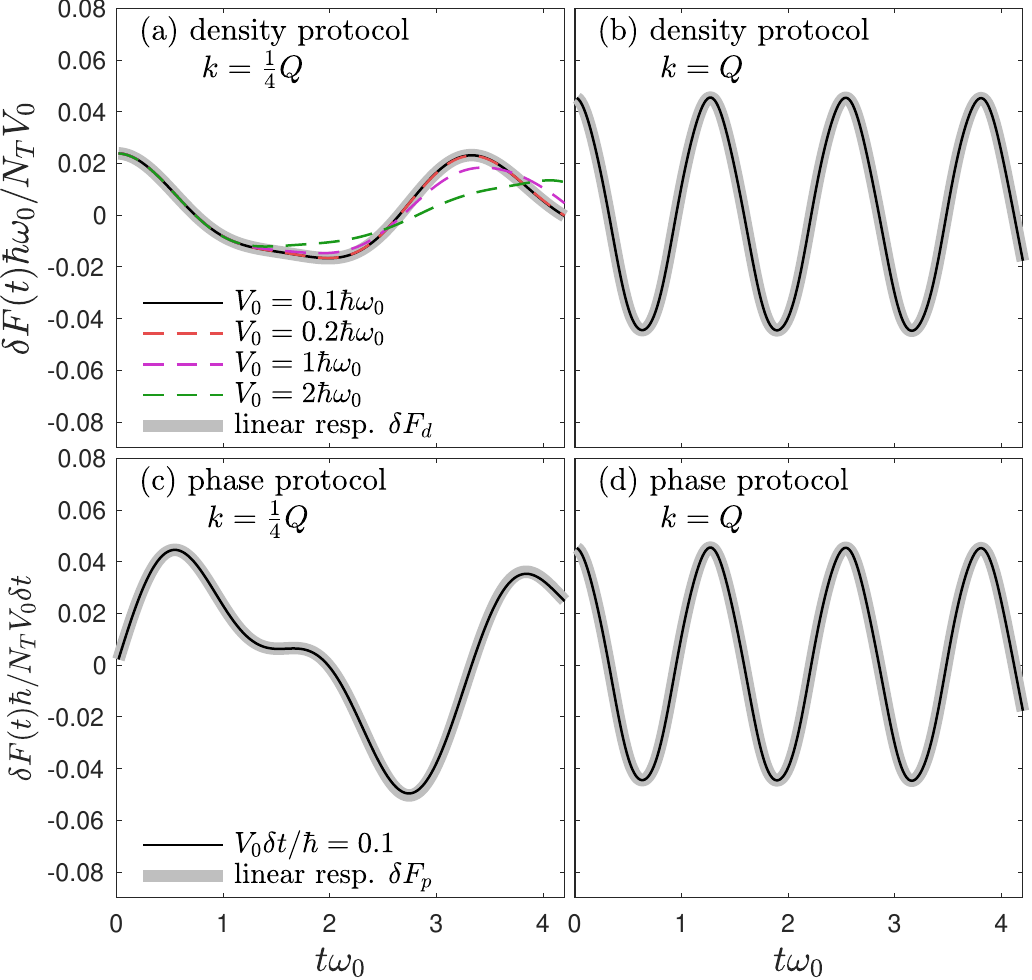} 
\caption{Evolution of the observable $\delta F(t)$ following the spectroscopy protocols. (a), (b) Density protocol and (c), (d) phase protocol results. 
GP results (lines)  and the linear response theory  (thick grey line) are shown. In (a)  the GP results are shown at various perturbation strengths to validate the linear response regime.  In (b) $V_0=0.1\hbar\omega$, while in (c) and (d) $V_0\delta t/\hbar=0.1$.
Ground state parameters: $\Lambda=25$,  $f_{s}=0.705$ [see Sec.~\ref{Sec:SF}] and $\mu=24.2\hbar\omega_{0}$, where   $\omega_{0}=\hbar/ma_{\mathrm{sc}}^{2}$. System length $L=12.2a_\mathrm{sc}$, supporting an $M_s=8$ site supersolid.
\label{figLRdynamics} }
\end{figure}

In Figs.~\ref{figLRdynamics}(c) and (d) we show  results for $\delta F(t)$ obtained from GP simulations following the phase protocol. 
 Similar to the observations for the density protocol case, we observe that probing with a low $k$ value [Fig.~\ref{figLRdynamics}(c)] excites a response with
two dominant frequency components, whereas at $k=Q$  [Fig.~\ref{figLRdynamics}(d)]  the response has a single dominant frequency.

\begin{figure}[htbp]  \includegraphics[width=3.2in]{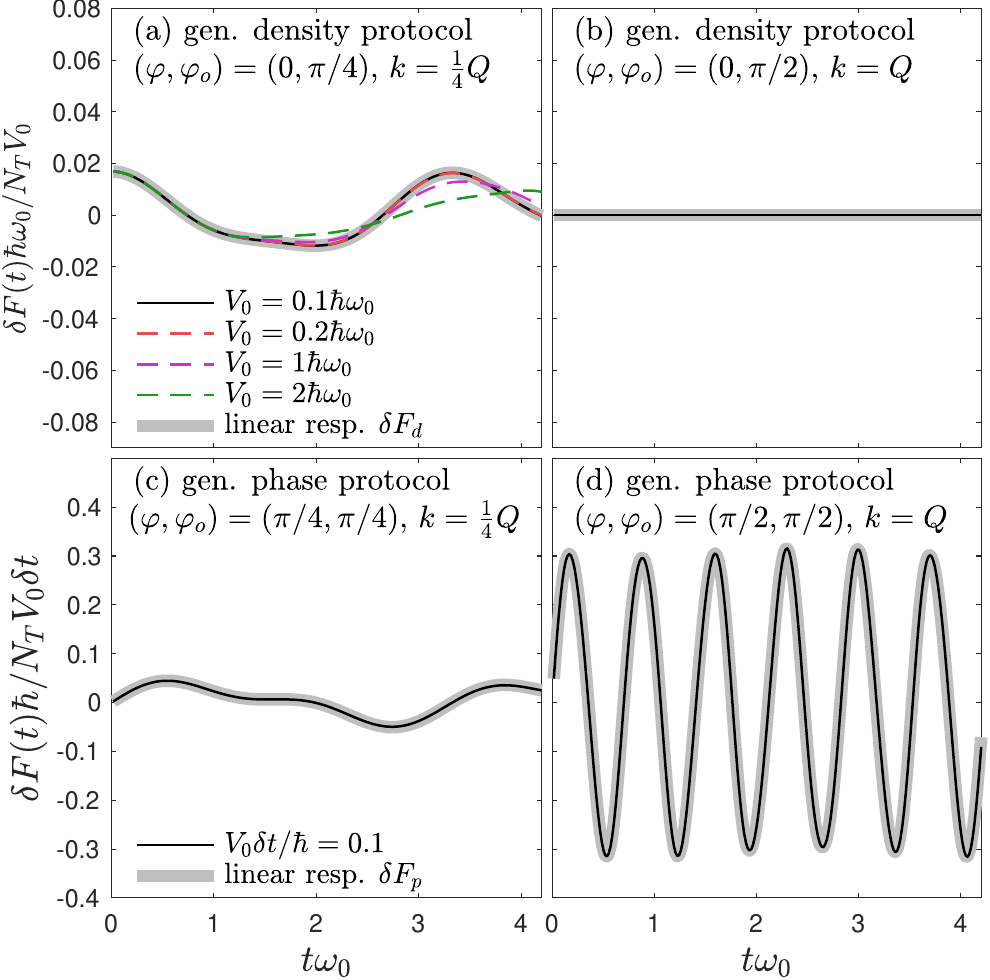} 
\caption{Evolution of the observable $\delta F(t)$ following the generalized spectroscopy protocols. (a), (b) Generalized density protocol and (c), (d) generalized phase protocol results. 
GP results (lines)  and the linear response theory  (thick grey line) are shown. In (a) the GP results are shown at various perturbation strengths to validate the linear response regime. Other parameters as in Fig.~\ref{figLRdynamics}.
\label{figLRgendynamics}  }
\end{figure}

\subsection{General density and phase protocol} \label{Sec:GenP}
The previously introduced protocols have the perturbation and observable aligned to the supersolid (see discussion as the end of  Sec.~\ref{Sec:SysPert}). This choice is most significant for $k=Q$ where the perturbation potential peaks and troughs all occur at the density peaks of the supersolid.

This motivates us to generalize the protocols to examine the effect of offsets of the perturbation and observable, relative to the supersolid.  We can generalise the perturbation potential for the phase protocol to
\begin{align} 
V_{p}(x,t) & = -V_{0}\delta t\,\delta(t)\cos(kx-\varphi),
\end{align}
where $\varphi$ is the phase offset of the perturbation, relative to the supersolid. Such an adjustment for the density protocol is redundant, because the supersolid ground state translates so that a lattice site (density peak) aligns with a potential minima (i.e.~effectively returning to the $\varphi=0$ case  of  $V_{d}$). However, for both protocols we can choose an arbitrary phase relative to the supersolid ($\varphi_{o}$) for the observable, i.e.~
\begin{align}
{F}(t)=\int dx\,\cos(kx-\varphi_o)|\psi(x,t)|^2.
\end{align}

Some results for the general spectroscopy protocols are shown in Fig.~\ref{figLRgendynamics}. The general density protocol results in Figs.~\ref{figLRgendynamics}(a) and (b) only  differ from those in Figs.~\ref{figLRdynamics}(a) and (b) by the  observable phase $\varphi_o$. In these results $\delta F(t)$  is suppressed by a factor of  $\cos\varphi_o$ relative to the respective earlier $\varphi_o=0$ results. Notably, for Fig.~\ref{figLRgendynamics}(b), while the system is excited by the perturbation (i.e., same dynamics shown in Fig.~\ref{figLRdynamics}(b)], the $\pi/2$-displaced observable is insensitive to these dynamics, yielding $\delta F(t)=0$.

The general phase protocol results in Figs.~\ref{figLRgendynamics}(c) and (d) can differ  in both phases. For  $\varphi=\varphi_o$ and $k<Q$, the response obtained from the general phase protocol is identical to the original phase protocol [cf.~Figs.~\ref{figLRdynamics}(c)  and \ref{figLRgendynamics}(c)]. For $k=Q$ we find that the response is sensitive to the phase choice. The case in Fig.~\ref{figLRgendynamics}(d) has a single frequency response, however, compared to Fig.~\ref{figLRdynamics}(d), the response is much stronger and at a higher frequency.  For $k=Q$  and $\varphi\ne\varphi_o$ (not shown) the response has two frequency components, being a combination of the results in Figs.~\ref{figLRdynamics}(d) and \ref{figLRgendynamics}(d).

\section{Linear response theory}\label{Sec:LRT}

Here we outline a linear response theory to describe the results obtained in the previous section, with additional details of the theory given in the Appendix. We begin by introducing the system excitations before presenting the linear response theory. Then we examine the nature of the excitations and the relevant dynamic structure factors to explain the general properties.

\subsection{Excitations}\label{Sec:excitations}
The unperturbed ($V_{\mathrm{pert}}=0$) system ground state $\psi_0$ satisfies the time-independent GP equation $\mathcal{L}\psi_0=\mu\psi_0$, where $\mu$ is the chemical potential and we take $\psi_0$ to be real. Note that $\int dx\,\psi_0^2=N_T$. For supersolid ground states (with $\Lambda>\Lambda_c$) the excitations have a Bloch wave form and can be labelled by quasimomentum $q$ in the  first Brillouin zone, $q\in(-Q,Q]$, and band index $\nu$. The excitation modes $\{ u_{\nu q}(x) ,v_{\nu q}(x)\}$ and respective energies $\{\hbar\omega_{\nu q}\}$, satisfy the Bogoliubov-de Gennes (BdG) equations
\begin{align}
\!\begin{bmatrix} 
      \mathcal{L}+X-\mu & -X \\
      X & -(\mathcal{L}+X-\mu) \\
   \end{bmatrix}\!\begin{bmatrix} u_{\nu q} \\ v_{\nu q}  \end{bmatrix} =\hbar\omega_{\nu q}\begin{bmatrix} u_{\nu q} \\ v_{\nu q}  \end{bmatrix},\label{BdGEqs}
\end{align}
where 
 $X$ is defined so that  
\begin{align}
Xf&= \psi_0(x)\int\!d {x}^\prime U_{\mathrm{sc}} ({x}-  {x}^\prime)f( {x}^\prime)\psi_0(x^\prime).  
\end{align}

\begin{figure}[htbp] \includegraphics[width=3.2in]{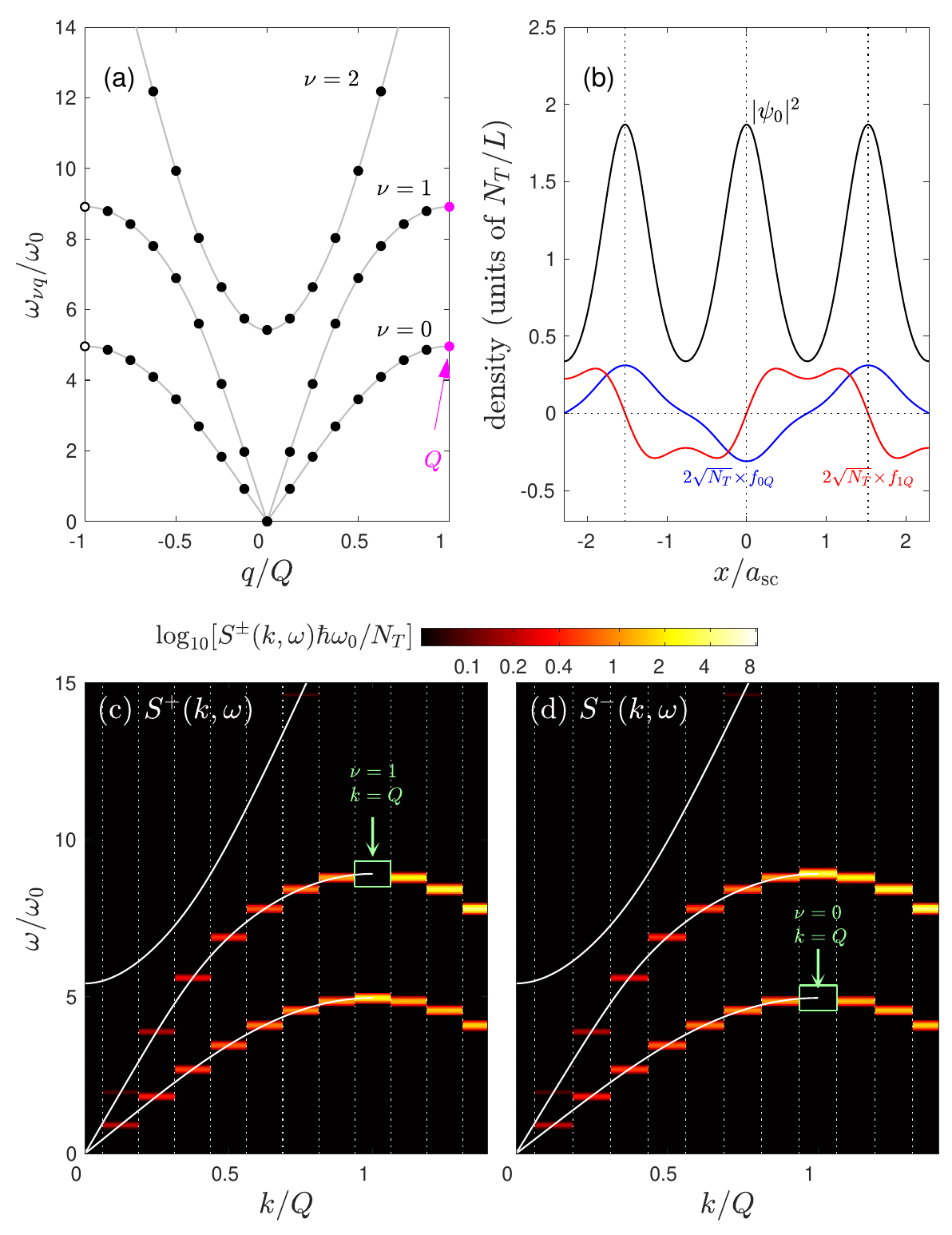} \caption{Excitations and dynamic structure factors for the translationally invariant ring supersolid. 
 (a) Excitation spectrum for a $M_s=16$ site supersolid (filled markers) and the infinite system limit (lines). The edge-modes with quasimomentum $Q$ are indicated. (b) Condensate density and the density fluctuations associated with the $\nu=0$ and $1$ edge-modes. The (c) $S^+$ and (d) $S^-$ the dynamic structure factors with the delta functions broadened to  $\delta(\omega)\to e^{-(\omega/\omega_{B})^{2}}/\sqrt{\pi}\omega_{B}$,
with $\omega_{B}=0.1\omega_{0}$.  Boxes and arrows indicate the edge state contribution to the dynamic structure factors. Other parameters as in Fig.~\ref{figLRdynamics}.  \label{figEdgeDSF}}
\end{figure}

We show results for the spectrum of a translationally invariant supersolid  in Fig.~\ref{figEdgeDSF}(a).   From the boundary conditions the excitation quasimomenta  are restricted to a discrete set determined by domain length $L$ and the number of supersolid sites\footnote{Allowed values are $q_n=\frac{2\pi n}{L}$, with $n\in \{-\frac{1}{2}M_s+1,\ldots, \frac{1}{2}M_s\}$ (cf.~allowed perturbation wavevectors described in Sec.~\ref{Sec:SysPert}). }. Continuous bands for the infinite system are shown  for reference, and help reveal the two gapless excitation bands where the energy of the excitations vanishes as $q\to0$. 
Labelling excitations by quasimomentum means that they are eigenstates of the translation operator. Because we have taken the supersolid to have a site at the origin, the system is also symmetric under the parity transformation. The intermediate states with $0<|q|<Q$  occur as degenerate pairs with $\pm q$ in each band and relate to each other by the parity operator. Exceptions are the state  $q=0$ and $q=Q$, which are unique in each band and are eigenstates of the parity operator. As a result the quasiparticles at $Q$ can be taken to be real and even or odd. We are not generally interested in $q=0$ excitations, although this case describes $\psi_0$, which can be  taken to be a real even solution.

It is useful to consider the density fluctuation associated with a quasiparticle. This can be defined by adding a quasiparticle to the condensate, i.e.~$\psi =\psi_0+u_{\nu q}e^{-i\omega_{\nu q}t}-v_{\nu q}^*e^{i\omega_{\nu q}t}$. To leading order (since the quasi particle amplitudes are a factor of $\sim\sqrt{N_T}$ smaller than the condensate wavefunction) the spatial density fluctuation is $|\psi|^2-\psi_0^2\approx 2\mathrm{Re}\{f_{\nu q}(x)e^{i\omega_{\nu q}t}\}$
where
\begin{align}
f_{\nu q}(x)=[u_{\nu q}^*(x)-v_{\nu q}^*(x)]\psi_{0}(x).
\end{align}
In general the $f_{\nu q}(x)$ are complex functions, however for the edge states $q=Q$ they are real and of definite parity (inherited from the symmetry of $\psi_0$ and $\{u_{\nu Q},v_{\nu Q}\}$ described above).  We show $f_{\nu Q}(x)$ for the lowest two bands in Fig.~\ref{figEdgeDSF}(b).  This reveals that the lowest band edge mode  $\{\nu=0,q=Q\}$ has a density fluctuation $f_{0 Q}(x)$ causing population exchange (i.e.~particle tunnelling) between adjacent sites. In contrast the first excited edge mode  $f_{1 Q}(x)$ causes the lattice sites to displace, with adjacent sites displacing in opposite directions.

\subsection{General linear response theory}
Linear response theory is well-established for a density coupled probe (e.g.~see \cite{BECbook}). A summary of this theory, specialized to the general potential considered here, is presented in the Appendix. This resulting prediction for the observable evolution is
\begin{align}
\delta F_d(t) & =\frac{V_0}{2}\sum_{\nu}\chi_{\nu}(k)\cos(\omega_{\nu \bar{k}}t),\label{dFd}\\
\delta F_p(t) & =\frac{V_0\delta t}{2}\sum_{\nu}\omega_{\nu \bar{k}}\chi_{\nu}(k)\sin(\omega_{\nu \bar{k}}t),\label{dFp}
\end{align}
for the general density and phase protocols, respectively\footnote{These two results are related because the two perturbations used in these protocols are related as $V_p=-\delta t\frac{\partial }{\partial t}V_d$.}.
Here $\bar{k}$ denotes $k$ reduced to the first Brillouin zone by an integer number of reciprocal lattice vectors. We have introduced
\begin{align}
\chi_{\nu}(k)\equiv \chi^+_{\nu}(k)\cos\varphi\cos\varphi_{o}+\chi^-_{\nu}(k)\sin\varphi\sin\varphi_{o},\label{chinu}
\end{align}
being the $\nu$-band contribution to the generalized static response function, where   \begin{align}
\chi^\pm_{\nu}(k)&= \sum_{q}\frac{|\langle\nu,q| \delta\hat{\rho}_k^\dagger\pm\delta\hat{\rho}_k  |0\rangle|^2}{\hbar\omega_{\nu q}},\label{chipm}
\end{align} 
are the two quadrature components of the density fluctuation operator, $\delta \hat{\rho}_k$ (see Appendix). Here $|0\rangle$ denotes the quasiparticle vacuum state (i.e.~ground state) and $|\nu,q\rangle$ denotes a state with a single $\{\nu,q\}$-quasiparticle excited.
Only excitations with $q=\pm \bar{k}$ contribute to $\chi^\pm_{\nu}(k)$, and we can evaluate these matrix elements as
\begin{align}
    \chi_{\nu}^{\pm}(k)=\frac{2|\delta\rho_{k,\nu \bar{k}}|^2}{\hbar\omega_{\nu\bar{k}}}\begin{cases}
       1, &0<|\bar{k}|<Q,\\
       1\pm(-1)^\nu, &\bar{k}=Q,
   \end{cases}
 \label{chipmnu}
\end{align} 
where
\begin{align}
\delta\rho_{k,\nu q}&\equiv\int dx\,e^{ikx}f_{\nu q}(x).
\end{align} 
 
 In Figs.~\ref{figLRdynamics}   and \ref{figLRgendynamics}  the linear response results (\ref{dFd}) and (\ref{dFp})  are shown for comparison to the GP results. Note that $\varphi=\varphi_o=0$ for the results in Fig.~\ref{figLRdynamics}   so that $\chi_\nu(k)\to\chi_\nu^+(k)$.  
 An interesting feature  is the distinct behavior of the edge mode contribution to  Eq.~(\ref{chipmnu}). Notably, $\chi_{\nu}^{+}$  ($\chi_{\nu}^{-}$) is zero for odd (even) bands at $\bar{k}=Q$. In general the strong response comes from the lowest two bands and thus to a good approximation $\chi_{\nu}^{+}(Q)$ is determined by the edge mode of the ground band, whereas $\chi_{\nu}^{-}(Q)$ is determined by the edge mode of the first excited band. This explains the single frequency response observed in Figs.~\ref{figLRdynamics}(b), (d)   and \ref{figLRgendynamics}(d). We can also understand this result from the symmetry of the excitations. For example,   $\chi_{\nu}^{+}(Q)$   describes the coupling of the even-symmetry condensate orbital via the even-symmetry potential $\cos(kx)$ to the $\nu$-band excitation at $q=Q$. Since the $\nu=1$ band edge excitation is odd,  this matrix element vanishes.

\subsection{Dynamic structure factors}
For the cases where the excitation and observable are described by the same operator it is convenient to define a dynamic structure factor (see \cite{BECbook}). Here we do this for the two quadrature cases of the density fluctuation operator, i.e.~$\delta\hat{\rho}_k^\dagger\pm\delta\hat{\rho}_k$:
\begin{align}
S^{\pm}(k,\omega)&=\sum_{\nu,q}{|\langle\nu,q| \delta\hat{\rho}_k^\dagger\pm\delta\hat{\rho}_k  |0\rangle|^2}\delta(\hbar\omega-\hbar\omega_{\nu q}),\label{Spm}\\
&= \sum_\nu\hbar\omega_{\nu \bar{k}}\chi^{\pm}_\nu(k)\delta(\hbar\omega-\hbar\omega_{\nu \bar{k}}),\label{Spm2}
\end{align}
utilizing expression (\ref{chipm}) for the matrix elements. 

We show results for the $S^{\pm}(k,\omega)$ dynamic structure factors in Figs.~\ref{figEdgeDSF}(c) and (d). These results indicate the strength of coupling to the various bands for different perturbation wavevectors $k$. Notably, we see that the majority of the weight resides in the lowest two bands, such that it is a reasonable approximation to truncate the sum over $\nu$ in Eqs.~(\ref{dFd}) and (\ref{dFp}) to $\nu=0$ and $1$. These results also reveal the selective coupling of the edge states. Notably the $\{\nu=1,Q\}$-excitation vanishes in $S^{+}(k,\omega)$ and the $\{\nu=0,Q\}$-excitation vanishes in $S^{-}(k,\omega)$.

\subsection{Low $k$ behavior} 
For  $0<|k|<Q$, from result (\ref{chipmnu}), we have that 
\begin{align}
\chi_\nu^{\pm}(k)\to\chi_\nu^\rho(k)=\frac{2|\delta\rho_{k,\nu k}|^2}{\hbar\omega_{\nu k}},
\end{align}
 where $\chi_\nu^\rho(k)$ is the $\nu$-band contribution to the usual static density response function. The \ifmmode \check{S}\else Š\fi{}indik \textit{et
al.}~\cite{Sindik2024a} probing scheme was proposed for the long wavelength limit (and for  $\varphi=\varphi_o$), such that  (\ref{dFd}) reduces to their result
\begin{align}
\delta F_d(t) & =\frac{V_{0}}{2}\sum_{\nu=0,1}\chi_{\nu}^\rho(k)\cos(\omega_{\nu  {k}}t),\quad 0<|k|<Q,\label{dFdS}
\end{align}
relating directly to the density response function.  
Applying similar arguments to the general phase protocol with $\varphi=\varphi_o$, allows us to write the small $k$ linear response in terms of the static density response function as
\begin{align}
\delta F_p(t) & =\frac{V_{0}\delta t}{2}\sum_{\nu=0,1}\omega_{\nu {k}}\chi_{\nu}^\rho(k)\sin(\omega_{\nu {k}}t),\quad 0<|k|<Q.\label{dFpS}
\end{align}

\begin{figure}[htbp] \includegraphics[width=3.2in]{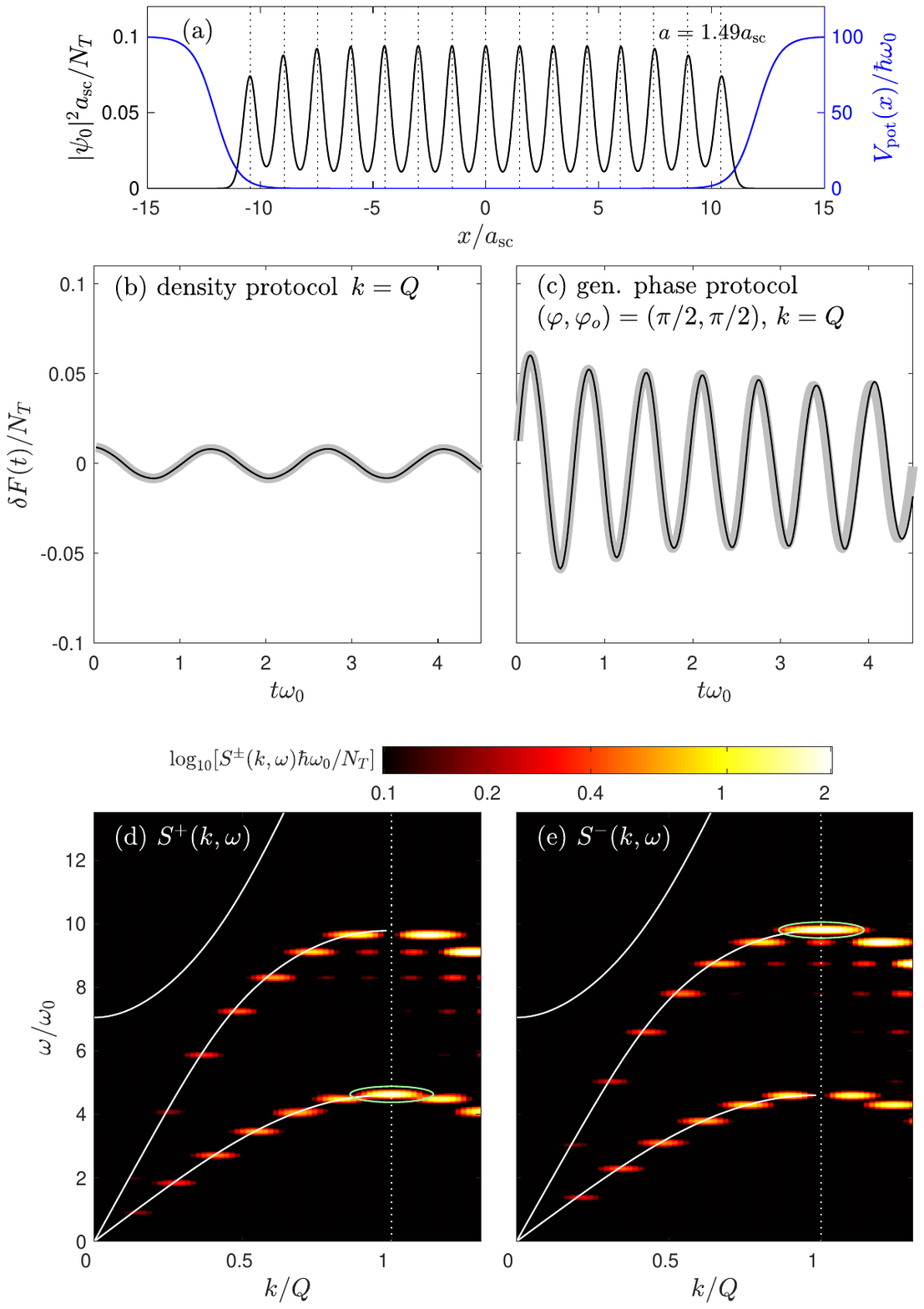} \caption{Spectroscopy protocols applied to a box trapped supersolid. (a) Ground state density and trapping potential. Examples of (b) density and (c) generalized phase  protocol responses for $V_0=0.2\hbar\omega_0$ and $V_{0}\delta t/\hbar=0.2$, $\varphi=\varphi_o=\pi/2$, respectively. GP dynamics (black line) and linear response theory (thick grey line). (d) $S^+$ and (e) $S^-$ the dynamic structure factors, frequency broadened as described in Fig.~\ref{figEdgeDSF}. Vertical dotted line indicates $k=Q$ and green ellipse indicates dominant feature at this wavevector.
Excitation spectrum of translationally invariant case with $\Lambda=27.3$ shown for comparison.  (b,c) Black line is from GP simulation and grey thick line is the linear
response result determined from the BdG excitations.     Results for a box trap potential $V_{\mathrm{pot}}=50\hbar\omega_0[\tanh\left(|x/a_{\mathrm{sc}}|-12\right)+1]$, with $N_TU_{0}=300\hbar\omega_{0}$.   \label{figboxtrap}}
\end{figure}

\begin{figure}[htbp] \includegraphics[width=3.2in]{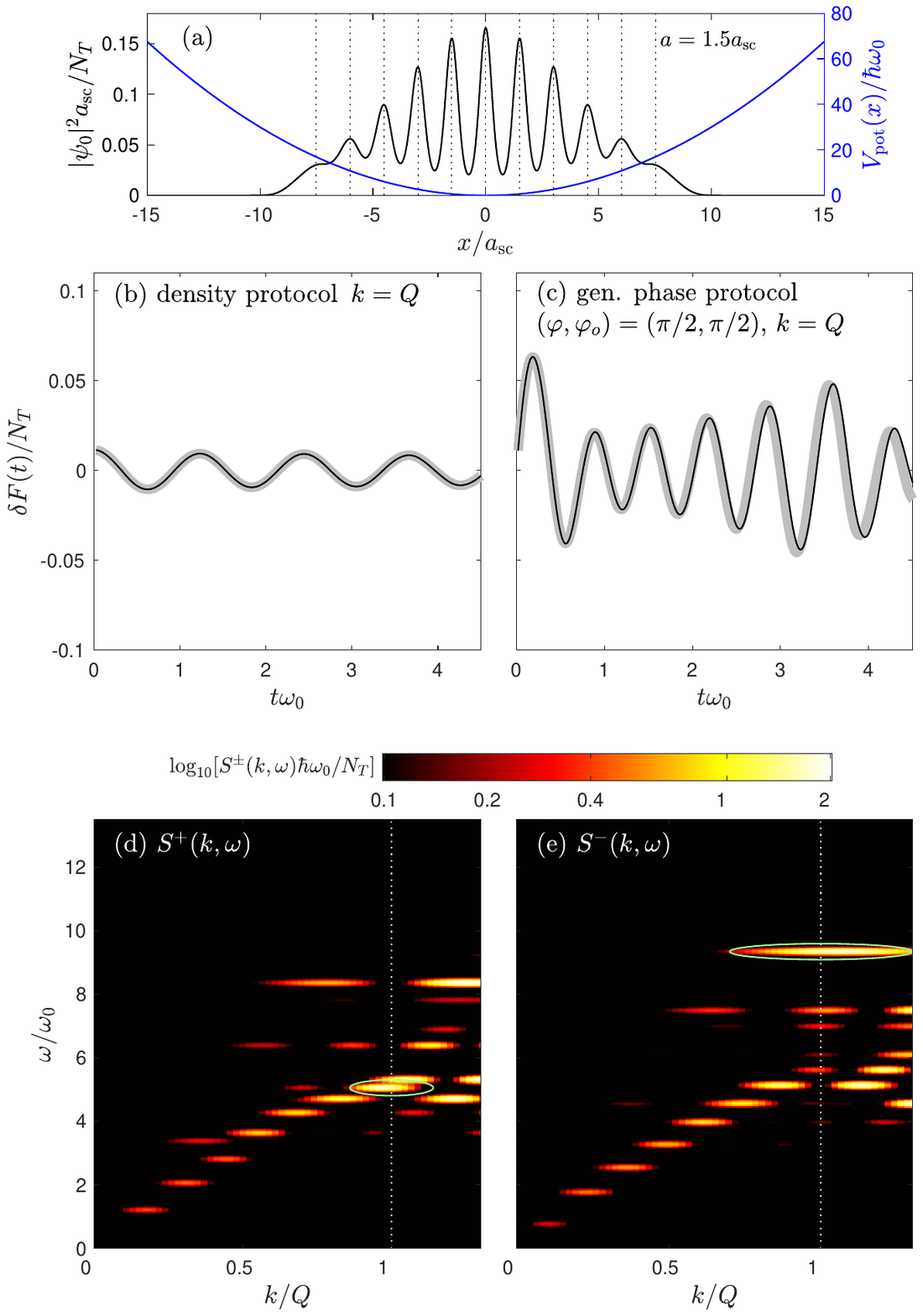} \caption{Spectroscopy protocols applied to a harmonically trapped supersolid. (a) Ground state density and trapping potential. Examples of (b) density and (c) generalized phase  protocol responses for $V_0=0.2\hbar\omega_0$ and $V_{0}\delta t/\hbar=0.2$, $\varphi=\varphi_o=\pi/2$, respectively. (d) $S^+$ and (b) $S^-$ the dynamic structure factors, frequency broadened as described in Fig.~\ref{figEdgeDSF}. Vertical dotted line indicates $k=Q$ and green ellipse indicates dominant feature at this wavevector.  
(c,d) Black line is from GP simulation and grey thick line is the linear
response result determined from BdG calculations. Inset to (c) shows
the ground state density profile.   Results for the harmonic trap $V_{\mathrm{pot}}=0.3(x/a_{\mathrm{sc}})^2\hbar\omega_0$, with $N_TU_{0}=175\hbar\omega_{0}$.   \label{figharmtrap}}
\end{figure}

\section{Extension to trapped cases}\label{Sec:Trap}

It is of interest to explore the application of the spectroscopy protocols to trapped
cases where translational invariance is broken. In this section we consider two types of trapped systems that could be explored in experiments:  a box-shaped trap and a harmonic trap.  The results for these two systems are presented in Figs.~\ref{figboxtrap} and \ref{figharmtrap}. In both cases subplot (a) shows the ground state density and the trapping potential for reference. The lattice sites are not strictly equally spaced in the presence of an external potential, but in both cases the peak spacing is well characterized by an average lattice constant $a$ (with corresponding lattice sites indicated by vertical dotted lines). We use  $a$ to define the reciprocal lattice vector for the spectroscopy protocols. Here we choose to focus on band edge probing, i.e.,~Eqs.~(\ref{VpDensity})  and (\ref{VpPhase}) with $k=Q$.

Results of the GP simulations of the dynamics are shown in subplots (b) and (c). This is seen to be in good agreement with the linear response theory. Because these systems are not translationally invariant, $\nu$ and $q$ are not good quantum numbers, and the response is determined by summing over all excitation modes [see  Eqs.~(\ref{dFdj}) to (\ref{chipmj}), which generalize the linear response theory of Eqs.~(\ref{dFd}), (\ref{dFp}) and (\ref{chipmnu})].

The box-trapped case has a relatively uniform average density  [see Fig.~\ref{figboxtrap}(a)] and the dynamic structure factors reveal a clear band structure comparable to the translationally invariant results [cf.~Figs.~\ref{figboxtrap}(d) and (e) and Figs.~\ref{figEdgeDSF}(c) and (d)].  To make a more direct comparison we can map the parameters to a similar translationally invariant case: the length of the box trapped state is  $L\approx 22a_{\mathrm{sc}}$, giving a dimensionless interaction parameter $\Lambda\approx27.3$ [from Eq.~(\ref{Lambda})]. The corresponding excitation bands for the infinite translationally invariant system at this value of $\Lambda$ are shown in Figs.~\ref{figboxtrap}(d) and (e) and seen to be in good quantitative agreement with the dynamic structure factor. We see the selective edge-mode behavior in these results. Notably, a mode of the lowest band  at $k\approx Q$ is seen to contribute strongly to $S^+$ but is absent from $S^-$ [indicated by ellipse in  Fig.~\ref{figboxtrap}(d)], and a mode of the first excited band at $k\approx Q$ is seen to contribute strongly to $S^-$ but is absent from $S^+$  [indicated by ellipse in  Fig.~\ref{figboxtrap}(e)].

In the harmonically trapped system the average density varies across the sample, although there is still a reasonably well-defined average lattice constant  [see Fig.~\ref{figharmtrap}(a)]. Here the response function does not reveal two clearly defined low energy bands like in the box-trapped case [Fig.~\ref{figharmtrap}(d) and (e)]. Although at the band-edge we again see a strong contribution from a low energy mode to $S^+$ that is absent from $S^-$  [indicated by ellipse in  Fig.~\ref{figharmtrap}(d)]  and a strong contribution from a higher energy mode to $S^-$ that is absent from $S^+$  [indicated by ellipse in  Fig.~\ref{figharmtrap}(e)]. However, the presence of other weaker modes with weight at the band edge is less clear, particularly for the case sensitive to probing an observable  related to $S^-$ [i.e.~Fig.~\ref{figharmtrap}(c)] where a beating between several frequencies is apparent.

\section{Relationship to superfluidity}\label{Sec:SF}
 
 The spectroscopy proposal by \ifmmode \check{S}\else Š\fi{}indik \textit{et
al.}~\cite{Sindik2024a}  and the spectroscopy experiment by Biagioni \textit{et al.}~\cite{Biagioni2024a} were applied to determine the superfluid fraction. This is of interest because the superfluid fraction of a supersolid at zero temperature is reduced from unity, even when the condensate fraction is unity. In general the superfluid fraction is determined by examining the response (current or energy) of the system to a small imposed phase gradient (i.e.,~imposed phase twist $\Delta \theta$ over the length of the system or equivalently a superfluid velocity $v_s=\hbar\Delta\theta/mL$) 
\ \cite{Leggett1970a,Sepulveda2010a,Roccuzzo2019a,Blakie2024a}. For example, by analysing the energy response, we have that the superfluid fraction is 
\begin{align}
f_s=\frac{1}{mN_T}\frac{\partial^2 {E}}{\partial v_s^2},\label{fsE}
\end{align}
where $E$ is the  energy functional.
Leggett developed an upper bound for the superfluid fraction in terms of the system density profile \cite{Leggett1970a,Sepulveda2008a} 
\begin{align}
f_s^+=\frac{L}{\rho}\left(\int dx\,\frac{1}{|\psi_0(x)|^2}\right)^{-1},\label{fsLegg}
\end{align}
where $\rho=N_T/L$ is the average density. 
 This bound is exact for the 1D soft-core model, and is an accurate estimate for 1D dipolar supersolids (e.g.~see results in Ref.~\cite{Smith2023a}).

Recently two experiments determined the superfluid fraction of a BEC in an optical lattice \cite{Chauveau2023a,Tao2023a} validating the Leggett bound. In this case the superfluid fraction is related to the speed of sound (in the optical lattice) $c$ as
\begin{align}
f_s=\frac{c^2}{c_\kappa^2},\label{fslatt}
\end{align}
where $mc_{\kappa}^2\equiv(\rho\kappa)^{-1}$, with $\kappa$ being the compressibility.  
For a supersolid, the spontaneously broken translational symmetry leads to the emergence of another gapless excitation band. A  1D supersolid exhibits two speeds of sound \cite{Watanabe2012a,Roccuzzo2019a,Blakie2023a,Ilg2023a,Platt2024a} and the simple result (\ref{fslatt}) no longer holds.

\subsection{Supersolid hydrodynamics: Long wavelength spectroscopy}

The hydrodynamic theory for Galilean invariant supersolids
(e.g.~see \cite{Andreev1969a,Salsow1977a,Son2005a,Josserand2007b,Yoo2010a,Hofmann2021a,Platt2024a}) furnishes a relationship between the superfluid fraction and the speeds of sound for a  supersolid:
\begin{align}
f_{s}=\frac{c_{1}^{2}c_{0}^{2}}{c_{\kappa}^{2}(c_{1}^{2}+c_{0}^{2}-c_{\kappa}^{2})}.
\end{align}
Here $c_0$ and $c_1$ are the speeds of sound of the lowest two (longitudinal) gapless excitation bands.

\ifmmode \check{S}\else Š\fi{}indik \textit{et
al.}~\cite{Sindik2024a} showed that performing density spectroscopy for $|k|\ll Q$ can determine the quantities in this expression for $f_s$.  Notably,
 measuring the response and fitting the results to Eq.~(\ref{dFdS}) determines $\omega_{\nu k}$ and $\chi^\rho_{\nu}(k)$ for $\nu=0,1$ [cf.~Fig.~\ref{figLRdynamics}(a) as an example of spectroscopy in this regime]. This information
gives the speeds of sound for the lowest two branches,
i.e.~$c_{\nu}=\lim_{k\to0}\omega^\rho_{\nu k}/k$ and  the compressibility $\kappa=\rho^{-1}\lim_{k\to0}\sum_{\nu}\chi^\rho_{\nu}(k)$.

This approach applies to the translationally invariant supersolid, and was specifically formulated for a dipolar supersolid in a ring trap. It has the disadvantage that the time scales of low-$k$ modes are slow, thus requiring long observation times to make the required fits. Furthermore, the density protocol requires a long initialisation step, i.e.,~waiting sufficiently long for the system to relax to the ground state of the perturbation before it is removed and the observable is measured.  Long time scales pose a challenge for dipolar supersolid experiments, where three-body loss tends to limit the lifetime. For this reason the phase protocol might be favorable for experiments, because it provides access to the same quantities [i.e.~by fitting the response to Eq.~(\ref{dFpS})], yet does not require the initialisation step.

\begin{figure}[htbp]  
 \includegraphics[width=3.2in]{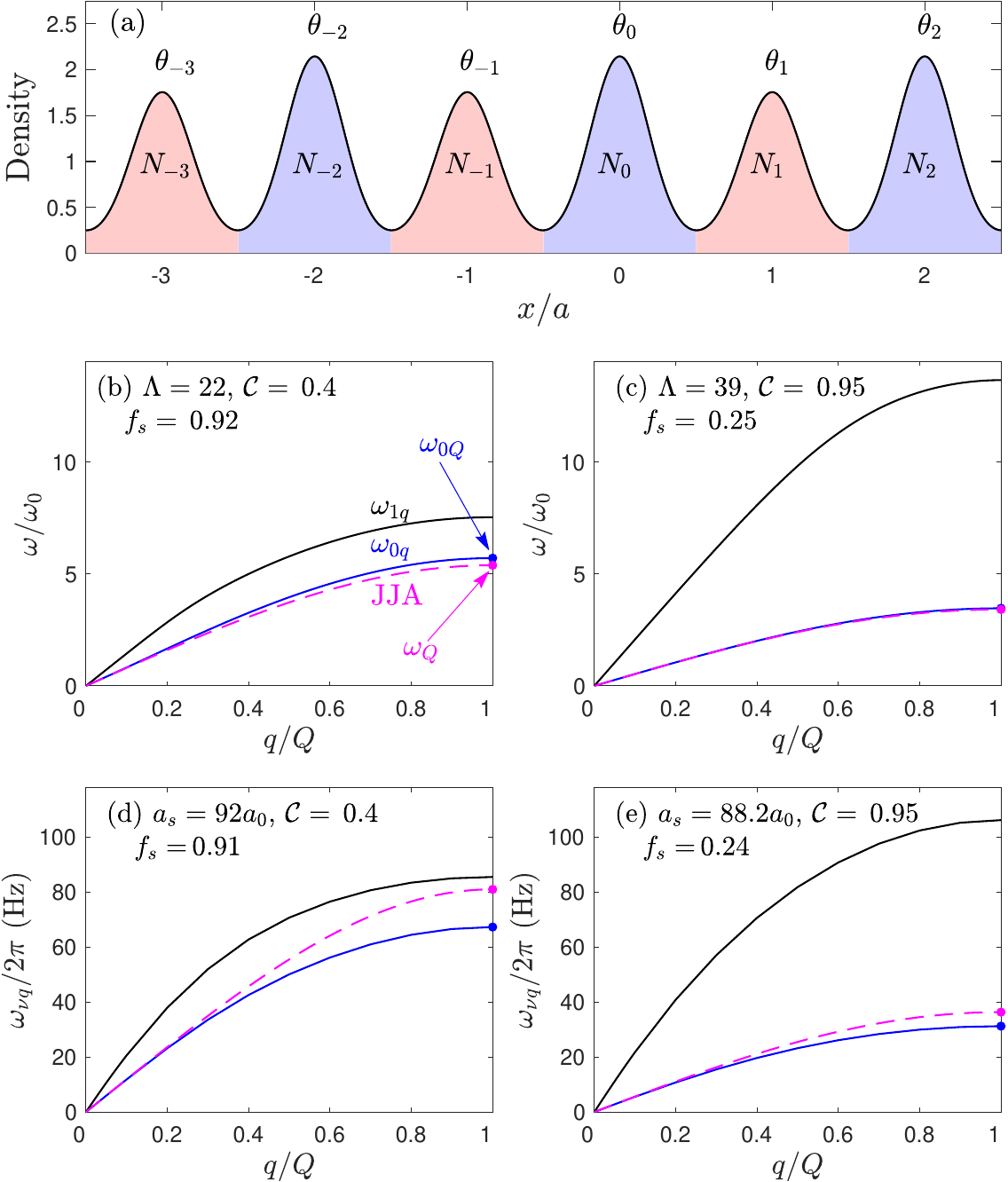}
  \caption{(a) Schematic of JJA model of a supersolid indicating the number $N_j$ and phase $\theta_j$ at site $j$. Here showing the case of an instantaneous population imbalance between even and odd sites, characteristic of the (band-edge) Josephson oscillation mid-cycle. Comparison of the JJA mode excitations (dashed magenta line) and BdG calculations
for the lowest sound excitation bands (blue and black lines) for (b), (c) the 1D soft-core
and (d), (e) tube dipolar supersolid states. Parameters indicated in subplots. The dipolar results are from the data set used in Refs.~\cite{Blakie2024a} to describe a $^{164}$Dy condensate of linear  density $\rho=2500\,\mu$m$^{-1}$ with radial confinement of $150\,$Hz. \label{figJJAcomp}}
\end{figure}

\subsection{Josephson-Junction array theory: Band-edge spectroscopy}\label{Sec:JJA}

A Josephson-Junction array (JJA) is a model for a BEC in an optical lattice  \cite{Anderson1998a,Cataliotti2001a} (also see \cite{Jaksch1998a}) and is an appealing model for a supersolid, where it can describe the coherent atom tunnelling dynamics between sites \cite{Ilzhofer2021a} (also see \cite{Buhler2023a}).  This model involves two parameters, the tunnel coupling between sites $J$  and the interaction parameter $U$, describing the interactions at each site. The system is then specified by the number of atoms $N_j$ at site $j$ and the phase of these atoms $\theta_j$ [see Fig.~\ref{figJJAcomp}(a)]. Within the JJA model the superfluid fraction is given by [from Eq.~(\ref{fsE})]
\begin{align}
f_{s}=\frac{Jma^{2}}{\hbar^{2}},\label{fsJJ}
\end{align}
and thus can be determined by measuring $J$.

Here we analyse the appropriateness of the JJA model for a translationally invariant supersolid where the ground state has $\bar{N}=N_T/M_s$ atoms at each site. We focus on the band-edge excitation of the system, where the disturbance alternates at adjacent sites, schematically shown as a density wave in  Fig.~\ref{figJJAcomp}(a). Following \cite{Biagioni2024a} we refer to the periodic oscillation dynamics of this state as being Josephson oscillation (cf.~DC Josephson effect for supersolids discussed in Ref.~\cite{Kunimi2011a}). Note, this is the kind of state and the dynamics occurring in Figs.~\ref{figLRdynamics}(b) and (d).  In this case all even sites are equivalent and all odd sites are equivalent, and we can study the dynamics in terms of the variables $\Delta N=N_1-N_0$ and $\Delta\theta=\theta_1-\theta_0$, being the atom number difference and phase difference between adjacent sites.  For a weak perturbation from equilibrium (i.e.,~cases where $|\Delta N|\ll \bar{N}$ and $|\Delta \theta|\ll1$) the dynamics of these quantities satisfies the Josephson-like equations \cite{Smerzi1997a}
\begin{align}
\hbar\Delta\dot{N} & =8\bar{N}J\Delta\theta,\\
\hbar\Delta\dot{\theta} & =-\frac{1}{\bar{N}}(2J+\bar{N}U)\Delta N,
\end{align}
with a harmonic solution of frequency  
\begin{align}
\omega_{Q}=\hbar^{-1}\sqrt{4J(4J+2\bar{N}U)}.\label{wQ}
\end{align}
Biagioni~\textit{et al.}~\cite{Biagioni2024a} used the  phase protocol at $k=Q$ to write a phase difference on adjacent sites of the supersolid, thus exciting  the Josephson oscillation [cf.~Fig.~\ref{figLRdynamics}(d) as an example of spectroscopy in this regime]. Measuring this frequency ($\omega_Q$) in the experiment, and with the additional input of $U$ from calculations, determines the value of $J$ [from Eq.~(\ref{wQ})], and hence the superfluid fraction $f_s$ via Eq.~(\ref{fsJJ}).

As noted in Refs.~\cite{Ilzhofer2021a,Buhler2023a}, the neglect of the crystal motion in the JJA, means this model is incomplete. Here we test its applicability by making a direct comparison of the JJA model to two supersolids: the 1D soft-core system we have discussed thus far in the paper, and a 1D dipolar supersolid (the physical system studied in Ref.~\cite{Biagioni2024a}).  We compare the excitations of the JJA model to those for the supersolid system obtained by numerical calculations of the BdG equations. The results for the dipolar system are from the data presented in Ref.~\cite{Blakie2023a} and we refer to that paper for the theoretical description of the system and calculation details.  While the JJA has a single gapless band the supersolids have two gapless bands [see Figs.~\ref{figJJAcomp}(b)-(e)]. The lowest band of these bands, known as second sound or the phase band, is the relevant band for comparison to the JJA result. This band is dominated by the tunnelling of atoms between sites, analogous to the physics described by the JJA.  The upper  band of the supersolid, known as first sound or the density band, is predominantly a crystal phonon-like excitation, i.e.,~involves a deformation of the supersolid crystal lattice.

The excitations of the translationally invariant JJA has the analytic form \cite{Rey2003a}
\begin{align}
\omega_{q}=\sqrt{\omega_{q}^{0}(\omega_{q}^{0}+2\bar{N}U/\hbar)},\label{BHBdG}
\end{align}
where $\hbar\omega_{q}^{0}=4J\sin^{2}({qa}/{2})$ and $q$ is the quasimomentum. 
To make the comparison it is necessary to determine the parameters $U$ and $J$. In deep optical lattices where this can be done using localized Wannier orbitals \cite{Jaksch1998a,Paul2016a,Blakie2004a}, however this approach is inapplicable to supersolids where there is significant overlap between sites. Here we identify $U$ and $J$ to reproduce the long-wavelength hydrodynamic properties of the supersolids. The two hydrodynamic properties we use are the superfluid fraction $f_s$ and the speed of sound of the lowest band $c_0$. These quantities are both determined from the numerical calculations of the BdG equations [Eq.~(\ref{BdGEqs}) and Ref.~\cite{Blakie2023a}] and ground state properties [Eq.~(\ref{fsE}) and Refs.~\cite{Smith2023a,Blakie2024a}]. From $f_s$ and $c_0$ the values of $J$ and $U$ in the corresponding JJA are thus determined:   $f_s$  gives  $J$ using Eq.~(\ref{fsJJ}), and subsequently  $c_0$  fixes the value of $U$ using the relationship
\begin{align}
c_{0}=\frac{a}{\hbar}\sqrt{2J\bar{N}U},
\end{align}
[from Eq.~(\ref{BHBdG})].  

Figure~\ref{figJJAcomp} presents comparisons for two cases of each for each supersolid system: (b,d) a relatively low contrast (high superfluid fraction) state and (c,e) a high contrast (low superfluid fraction) state. Here the contrast is defined as
\begin{align}
\mathcal{C}=\frac{\rho_{\max}-\rho_{\min}}{\rho_{\max}+\rho_{\min}},
\end{align} where $\rho_{\max}$ ($\rho_{\min}$) is the maximum (minimum) of the linear density, with $\mathcal{C}=0$ being the uniform superfluid state, and $\mathcal{C}=1$ being where the linear density goes to zero between sites.
The agreement between the JJA model dispersion relation $\omega_q$ and supersolid lowest band  $\omega_{0q}$ is assured for $q\to0$ because of our choice of parameters to match the hydrodynamic properties. The deviation for large $q$ thus reveals physics beyond the JJA model in the supersolids. Most importantly for the Biagioni \textit{et al.}~\cite{Biagioni2024a} scheme is the comparison of the frequency or band-edge mode, i.e.,~the Josephson oscillation frequency $\omega_Q$, as this is the quantity that they measure experimentally. This mode is beyond the hydrodynamic description (due to its short wavelength), and its relationship to the hydrodynamic properties, and particularly the superfluid fraction, relies on the appropriateness of the JJA model. We have indicated the edge modes for comparison in Figs.~\ref{figJJAcomp}(b)-(e). Notably, the relevant $\omega_{0Q}$ mode from the supersolid excitations. In general we find that agreement between $\omega_Q$ and $\omega_{0Q}$ is quite reasonable for the cases we have examined, although it is noticeably better for the soft-core supersolid.  We understand this as arising because the soft-core model tends to have a more rigid lattice than the dipolar supersolid, as revealed by studies of the  elastic properties of these supersolids \cite{Platt2024a}. For the dipolar supersolid case of Fig.~\ref{figJJAcomp}(d) the relative difference between $\omega_Q$ and $\omega_{0Q}$ is about 20\%, and this would be reflected in an error in the inferred value of $f_s$.

\section{Conclusion}\label{Sec:Conclusions}
In this work, we introduced a linear response description linking and generalizing the density and phase protocols presented separately in recent works. We illustrate the theory using a soft-core model of a supersolid, but the theory is more generally applicable. For a translationally invariant system both protocols tend to excite excitations from two lowest gapless excitation bands with a wavevector set by the perturbation. 
Interestingly, our theory explains the peculiar behavior observed at the band edge—where only a single excitation responds at a wavelength twice the lattice constant. This phenomenon arises due to the symmetry of the edge modes, which can be selectively excited depending on the alignment of the external potential with respect to the supersolid crystal, i.e., via the generalized probing scheme we suggest.
 
 We present results for trapped cases, where the translational invariance is broken, finding the band-edge feature of excitations still approximately holds. Finally, we have discussed the superfluid fraction of a supersolid, and how this relates to hydrodynamic theory and a Josephson-Junction array model. These results provide valuable insights into the use of spectroscopy protocols for determining the superfluid fraction and deepen our understanding of supersolid excitations.

\section*{Acknowledgments}
\noindent  The authors acknowledge M. Cui and W. Cresswell for early work exploring Josephson dynamics in a supersolid that informed this study and funding from the Marsden Fund of the Royal Society of New Zealand. 
\appendix

\section*{Appendix: Linear response theory}\label{App:densityflucts} 
Within the framework of Bogoliubov
theory the field operator can be expressed as 
\begin{align}
\hat{\psi}(x)=\psi_{0}(x)+\sum_{j}[u_{j}(x)\hat{\alpha}_{j}-v_{j}^*(x)\hat{\alpha}_{j}^{\dagger}],
\end{align}
where  
 $\{\hat{\alpha}_{j},\hat{\alpha}_{j}^{\dagger}\}$ are bosonic
mode operators which
satisfy the commutation relations $[\hat{\alpha}_{i},\hat{\alpha}_{j}^{\dagger}]=\delta_{ij}$.
Here the excitations modes $\{u_{j}(x),v_{j}(x)\}$  (with respective energies $\{\hbar\omega_j\}$) are the generalization of Eq.~(\ref{BdGEqs}) to allow for an external potential, such that quasimomentum is not a good quantum number, and we introduce the general index $j$.

The density fluctuation operator, $\delta\hat{\rho}_{k}^{\dagger}=\int dx\,e^{ikx}(\hat{\psi}^{\dagger}\hat{\psi}-\psi_{0}^{2})$,
to first order in the quasiparticle operators, is given by 
\begin{align}
\delta\hat{\rho}_k^\dag=\sum_j(\delta\rho_{k,j}\hat{\alpha}^\dag_j + \delta\rho_{-k,j}^*\hat{\alpha}_{j}),\label{rhohatk}
\end{align}
where we have defined the matrix element as 
\begin{align}
\delta\rho_{k,j}&\equiv\langle j|\delta\hat{\rho}_{k}^\dag|0\rangle\\
&=\int dx\,e^{ikx}[u_{j}(x)-v_{j}(x)]^*\psi_{0}(x),\label{dfmej}
\end{align}
with $|0\rangle$ being the quasiparticle vacuum state, and $|j\rangle=\hat{\alpha}_{j}^{\dagger}|0\rangle$ being a state with a single $j$-quasiparticle. The density fluctuation operator is useful because the perturbation potential can be written in second quantized form as
\begin{align}
    \hat{V}=-\frac{1}{2}V_0(t)(\delta\hat{\rho}_k^\dagger e^{-i\varphi}+\delta\hat{\rho}_k e^{i\varphi}),
\end{align}
where $V_0(t)=V_0\theta_H(-t)$ for the generalized density protocol  or $V_0(t) = V_0\delta t\delta(t)$ for the phase protocol. 
Using time-dependent perturbation theory we obtain the following expressions for the response evolution
\begin{align}
    \delta F_d(t) &=  V_0\sum_j \frac1{4\hbar\omega_j}[c_j(k)e^{-i\omega_jt} + c_j^*(k)e^{i\omega_jt}],\\
      \delta F_p(t) & =  V_0\delta t\sum_j \frac{i}{4\hbar}[c_j(k)e^{-i\omega_jt} - c_j^*(k)e^{i\omega_jt}],
    \end{align}
    for the density and phase protocols, respectively, where
    \begin{align}
    c_j(k) &= \delta\rho^*_{-k,j}\delta\rho_{k,j} e^{-i(\varphi+\varphi_o)} + \delta\rho^*_{k,j}\delta\rho_{-k,j} e^{i(\varphi+\varphi_o)}\notag\\
      &+ |\delta\rho_{k,j}|^2 e^{-i(\varphi-\varphi_o)}+ |\delta\rho_{-k,j}|^2 e^{i(\varphi-\varphi_o)}.
\end{align}

For the trapped supersolid, $u_j-v_j$ can be taken to be real, so that $\delta\rho_{k,j}=\delta\rho_{-k,j}^*$, and either even $(\delta\rho_{k,j}=\delta\rho_{-k,j})$  giving $c_j(k) = 
4|\delta\rho_{k,j}|^2 \cos\varphi\cos\varphi_o$ or odd $(\delta\rho_{k,j}=-\delta\rho_{-k,j})$ giving $c_j(k)=4|\delta\rho_{k,j}|^2 \sin\varphi\sin\varphi_o$,  which can be written as
\begin{align}
    c_j(k) &= |\delta\rho_{k,j} + \delta\rho_{-k,j}|^2\cos\varphi\cos\varphi_o \notag\\
    &+|\delta\rho_{k,j} - \delta\rho_{-k,j}|^2\sin\varphi\sin\varphi_o.
\end{align}
Thus the previous results can be written in the form
\begin{align}
    \delta F_d(t) & = \frac{V_0}2\sum_j\chi_j(k)\cos(\omega_jt),\label{dFdj}\\
    \delta F_p(t) & = \frac{V_0\delta t}2\sum_{j}\omega_{j}\chi_{j}(k)\sin(\omega_{j}t),\label{dFpj}
\end{align} 
with
\begin{align}
\chi_j(k)\equiv \chi^+_{j}(k)\cos\varphi\cos\varphi_{o}+\chi^-_j(k)\sin\varphi\sin\varphi_{o},\label{chij}
\end{align}
being the $j$-band contribution to the generalized static response function, where   \begin{align}
\chi^\pm_j(k)&= \frac{|\langle j| \delta\hat{\rho}_k^\dagger\pm\delta\hat{\rho}_k  |0\rangle|^2}{\hbar\omega_j}.\label{chipmj}
\end{align}

For the translationally invariant supersolid the matrix elements become 
\begin{align}
\delta\rho_{k,j}\to \delta\rho_{k,\nu q}=\int dx\,e^{ikx}[u^*_{\nu q}(x)-v^*_{\nu q}(x)]\psi_{0}(x),
\end{align}
with the selection rule $k=q+2nQ$, where $n$ is an integer. Then, for $0<|q|<Q$,  $|\delta\rho_{-k,\nu-q}|^2=|\delta\rho_{k,\nu q}|^2$ and $\delta\rho_{k,\nu q} \ne 0 \implies \delta\rho_{-k,\nu q}=0$ so
\begin{align}
    c_{\nu q}(k) + c_{\nu -q}(k) &= 2(|\delta\rho_{k,\nu q}|^2 + |\delta\rho_{-k,\nu q}|^2) \cos(\varphi-\varphi_o). 
  \end{align}
Also,  $\delta\rho_{-k,\nu Q} = (-1)^\nu\delta\rho_{k,\nu Q}$, so
  \begin{align}
      c_{\nu Q}(k) &= 2|\delta\rho_{k,\nu Q}|^2[\cos(\varphi-\varphi_o) + (-1)^\nu\cos(\varphi+\varphi_o) ],
\end{align}
i.e.
\begin{align}
    \delta F_d(t) &= \frac{V_0}2\sum_{\nu} \chi_\nu(k)\cos(\omega_{\nu\bar k}t),\\ \delta F_p(t) &= \frac{V_0\delta  t}2\sum_{\nu} \omega_{\nu\bar k}\chi_\nu(k)\sin(\omega_{\nu\bar k}t),\end{align}
 with 
\begin{align}
    \chi_\nu(k)&=\frac{2|\delta\rho_{k,\nu\bar{k}}|^2}{\hbar\omega_{\nu\bar k}} 
        [\cos(\varphi-\varphi_o) 
        +(-1)^\nu\cos(\varphi+\varphi_o)\delta_{\bar{k},Q}]
    \end{align} 
which gives \eqref{chinu} using
\begin{align}
    \chi_\nu^\pm(k) &= \sum_q \frac{|\langle\nu,q|\delta\hat\rho_k^\dag \pm \delta\hat\rho_k|0\rangle|^2}{\hbar\omega_{\nu q}} = \sum_q \frac{|\delta\rho_{k,\nu q} \pm \delta\rho_{-k,\nu q}|^2}{\hbar\omega_{\nu q}}.
\end{align}


%

\end{document}